\documentclass{PoS}

\newcommand{\eq}{\begin{equation}}
\newcommand{\eqx}{\end{equation}}
\newcommand{\eqn}{\begin{eqnarray}}
\newcommand{\eqnx}{\end{eqnarray}}
\newcommand{\bi}{\begin{itemize}}
\newcommand{\ei}{\end{itemize}}
\newcommand{\nn}{\nonumber}
\newcommand{\ra}{\rangle}
\newcommand{\la}{\langle}
\newcommand{\bz}{\bar{z}}

\title{Beyond complex Langevin equations}

\ShortTitle{Beyond complex Langevin ...}

\author{\speaker{Jacek Wosiek}
        \\Jagellonian University\\
        E-mail: \email{Jacek.Wosiek@uj.edu.pl}}

\abstract{A simple integral relation between a complex weight and the corresponding positive distribution is derived by introducing a second complex variable. Together with the positivity and normalizability conditions, this sum rule allows to construct explicitly  equivalent pairs of distributions in simple cases. In particular the well known solution for a complex gaussian distribution is generalized to an arbitrary complex slope. This opens a possibility of positive representation of Feynman path integrals directly in the Minkowski time. Such construction is then explicitly carried through in the second part of this presentation. The continuum limit of the new representation exists only if some of the additional couplings tend to infinity and are tuned in a specific way. The approach is then successfully applied to three quantum mechanical examples including a particle in a constant magnetic field -- a simplest prototype of a Wilson line. Further generalizations are shortly discussed and an amusing interpretation of new variables is briefly mentioned. 
}

\FullConference{34th annual International Symposium on Lattice Field Theory\\
		24-30 July 2016\\
		University of Southampton, UK}

\begin{document}

\noindent {\em Basics and the main idea.}

Generally quantum averages result from weighting observables with complex amplitudes rather than with positive
probabilities. The technique colloquially referred to as Complex Langevin can in principle be used to replace this by a standard, statistical averaging over a suitably defined stochastic process. The method was proposed long time ago \cite{P,Kl}, but recently has attracted a new wave of interest, especially in studies of quantum chromodynamics at finite chemical potential \cite{S1,S2}. Still, contrary to the real Langevin approach, there is no general proof of the convergence \cite{S3,S4} and the evidence for the success is somewhat limited \cite{HW,Ph,Bl}. 

In this talk the recent construction of  the positive probabilities, without any reference to  stochastic processes and/or Fokker-Planck equations, is described \cite{Ja}. Thus above mentioned  difficulties are avoided albeit in a few simple test cases. 
Building on this the positive representation for three, classic Feynman path integrals directly in the Minkowski space is derived. 
Notably one of the discussed cases cannot be rendered positive by the Wick rotation.

The difficulties of Complex Langevin approach appear because of the poor convergence, if any, of complex stochastic processes employed so far in various constructions.
On the other hand we do not really need to generate the positive two dimensional distribution with the stochastic process in a complex plane.
The only and the real problem is to find a positive distribution $P(x,y)$ which reproduces averages over a complex density $\rho(x)$:  
$\la f(x) \ra_{\rho(x)}=\la f(x+i y) \ra_{P(x,y)}$.  To this end introduce the second, antiholomorphic variable $\bar{z}=x-i y$, and rewrite this "matching condition" in a more general form (all contours are chosen such that the integrals exist)
\eqn
 \frac{\int_{\Gamma_z} f(z) \rho(z) dz}{\int_{\Gamma_z} \rho(z) dz} = \frac{\int_{\Gamma_z} \int_{\Gamma_{\bz}} f(z) P(z,\bz) dz d\bz }{\int_{\Gamma_z} \int_{\Gamma_{\bz}} P(z,\bz) dz d\bz}. \label{B3}
\eqnx
This will be the case provided
\eqn
\rho(z)=\int_{\Gamma_{\bz}} P(z,\bar{z}) d \bar{z}.  \label{Pro}
\eqnx
This is the main result of the proposed "beyond Complex Langevin" approach. The connection between $\rho$ and $P$ turns out to be very simple if $z$ and $\bar{z}$ are considered as the two {\em independent} complex variables. The main challenge, however, remains in
finding $P(z,\bar{z})$ such that $P(z,z^*)=P(x,y)$ is positive and normalizable.
It is shown below that this program can in fact be carried through quantitatively, at least in few physically interesting cases, already providing 
some novel results.

\noindent {\em The generalized gaussian model}

A more general, than originaly derived in \cite{AY}, positive distribution can be obtained if we start from a generic quadratic action for $P(z,\bar{z})$ in two complex variables $z$ and $\bz$
\eqn
P(z,\bar{z})=e^{-S(z,\bar{z})},\;\;\;\;S(z,\bar{z})&=& a^* z^2 + 2 b z \bz + a \bz^2, \label{S2}
\eqnx
with an arbitrary complex $a=\alpha+i\beta$ and real $b=b^*$. Upon restriction to $\bar{z}=z^*$,  
$P(x,y)$ is positive and normalizable
provided $b > |a|$. 
At the same time the the $\bar{z}$ integral 
\eqn
\rho(z)=\int_{\Gamma_{\bz}} P(z,\bz) d \bz = 
  \frac{1}{2}\sqrt{\frac{\pi}{-a}}\exp{\left( - s z^2\right)},\;\;\;s=\frac{|a|^2-b^2}{a}. \label{ip2}
\eqnx
not only reproduces, the well known solution of \cite{AY}, but provides its generalization for all complex values of the slope $s$ \cite{Ja1}.
In particular consider two interesting special cases.

1.  For real and negative $s$, the complex density blows up along the real axis. On the other hand the distribution $P(x,y)$ is positive and normalizable at $\alpha>0$ and  $\beta=0$ producing the correct average over the "divergent" distribution $\rho$. This explains a ``striking example" observed in the literature \cite{AS}, namely that, upon change of variables, the complex Langevin simulation based on (\ref{ip2})  actually has the correct fixed point also for negative ${\cal R}e\;s$. The answer is that the positive distribution used until now is part of a richer structure (\ref{ip2}), which accommodates negative ${\cal R}e\;s$ as well.
 
2. Similarly, the complex density $\rho(z)$ for purely imaginary $s$ is readily represented by the positive distribution $P(x,y)$, which is perfectly well defined at $\alpha=0$ and arbitrary $\beta$, as long as $|\beta|<b$. This opens an exciting possibility of positive representations for Feynman path integrals directly in the Minkowski time. Such a construction is reviewed below.
 
 In both cases the original density of Ref.\cite{AY} does not exist.

\noindent {\em A nonlinear model }

Another possible solution can be derived if we start from the action
\eqn
S_4(z,\bz)=\frac{d}{b}(a^* z^2+2 b z \bz + a \bz^2)^2, \nn  
\eqnx
with complex $a$  and real $b/d > 0 $. The density $P(x,y)$ is again positive and normalizable on the $x,y$ plane.
The complex density $\rho_4(z)$ can be then obtained in a closed form as
\eqn
\rho_4(z)&=&\frac{i}{2}\int_{\Gamma_{\bz}} d \bz  e^{- S_4(z,\bz)}
=\frac{i}{2}\left(\frac{b}{2 d a^2}\right)^\frac{1}{4}  \exp{\left(-\sigma z^4\right) } \left(\sigma z^4 \right)^\frac{1}{4} 
K_{\frac{1}{4}}\left(\sigma z^4\right),\label{BK}
\eqnx
with an arbitrary complex
\eqn
\sigma=\frac{d (b^2-|a|^2)^2}{2 b a^2}.\nn 
\eqnx
All contours (here and below) are such that the integrals exists. Basically one can choose straight lines with slopes determined by 
the phase of $a$. 

The density (\ref{BK}) has a simple leading asymptotics 
\eqn
  \rho_4(z)\sim e^{\left(- 2 \sigma z^4 \right)} ,\;\;\; z \longrightarrow \infty, \nn
\eqnx
and therefore might be of some practical interest (e.g. in optimizing some reweighting algorithms). The main point of this example is however, that the original idea, namely constructing positive representations by introducing a second variable, seems to be general and points towards existence of some unexplored yet structures.

Obviously there is a lot of freedom in choosing an initial action. It remains to be seen to what extent this freedom allows to derive  complex densities of wider physical interest. 

\noindent {\em Many variables and quantum mechanical Minkowski path integrals}

For the action we take $N$ copies of (\ref{S2})  and add the nearest neighbour couplings, with periodic boundary conditions in $z_i$ and $\bz_i$: $z_{N+1}=z_1, \bz_{N+1}=\bz_1, z_{0}=z_N, \bz_{0}=\bz_N$, $a, c \in C$, $b \in R$,
\eqn
S_N(z,\bz)= \sum_{i=1}^N
 a \bz_i^2 + 2 b  \bz_i z_i + 2 c \bz_i z_{i+1} + 2 c ^* z_i \bz_{i+1} + a^* z_i^2,\;\;\;\;P_N(z,\bz)=e^{-S_N(z,\bz)} . \label{SN} 
\eqnx
The complex density $\rho(z)$ results from integrating $P_N(z,\bz)$ over all $\bz$ variables
\eqn
\rho(z)=\int \prod_{i=1}^{N}d\bz_i P(z,\bz)=\left(\frac{i}{2}\right)^N\int \prod_{i=1}^{N}d\bz_i \exp{\left(-S_N(z,\bz)\right)}\equiv
\exp\{-S^{\rho}_N\}.\nn
\eqnx
The integration is elementary and one obtains for the effective action ( $2c=2\gamma=-b + |a|$), 
\eqn
-S_N^{\rho}(z)= {\cal A} \sum_{i=1}^N    \left(   z_i -  z_{i+1} \right)^2
- r\left(z_{i-1}-z_{i+1}\right)^2,\;\;\;\;{\cal A}=\frac{b(b-|a|)}{a} ,\;\;r=\frac{b-|a|}{4b}.\;\;\;\;\label{Srho}
\label{rhN}
\eqnx
This is reminiscent, however does not quite agree, with the discretized Feynman action for a free particle. 
\eqn
S^{free}_N=\frac{i m}{2 \hbar\epsilon} \sum_{i=1}^N (z_{i+1}-z_i)^2,  \label{SFd}
\eqnx 
On the other hand, both constraints, namely ${\cal A}  \rightarrow \frac{i m}{2 \hbar\epsilon}$ ,  and $r= \rightarrow 0$,  
 can be satisfied in the limit  (referred from now on as $\lim_1$)
\eqn
|a|, b \rightarrow \infty, b-|a|=\frac{ m}{2 \hbar\epsilon}=const.\equiv d,\;\;a=-i |a| . \label{lim1}
\eqnx
This completes the construction of the positive representation for the path integral of a free particle directly in the Minkowski time.

All quantum averages can now be obtained by weighting suitable, i.e. complex in general, observables with the positive and normalizable distribution $P_N(x_i,y_i)=\exp{\left(-S_N(z_i,z_i^*)\right)}$
, and then taking  the limit (\ref{lim1}) followed by the continuum limit: $N\rightarrow\infty,\epsilon\rightarrow 0, N\epsilon=const\equiv T $. 

Upon slightly different identification of parameters the action (\ref{SN}) covers also the harmonic oscillator case (with the frequency $\omega$). 
\eqn
-S_N^{\rho}(z)=\frac{i m }{2 \hbar \epsilon}\left( (z_1-z_2)^2 - \frac{\omega^2 \epsilon^2}{2}(z_1^2+z_2^2)\right)+ (nnn) .  \nn
\eqnx
Similarly to the free particle, the $nnn$ terms will vanish for large $|a|$ and $b$. However the first limit ($lim_1$) has to be taken 
along little more complicated trajectory. A possible parametrization in terms of one independent variable $\nu\rightarrow 0$, is ($a=-i |a|$)
\eqn
b=\frac{\mu}{\nu},\;\;\; |a|= \frac{\mu}{\nu} \zeta(\nu,\rho),\;\;\; 2\gamma=-\mu \zeta(\nu,\rho),  \;\;\;
\zeta(\nu,\rho)= \frac{\sqrt{1-2\nu^2\rho+\nu^2\rho^2}-\nu(1-\rho)}{1-\nu^2},   \label{hotraj}
\eqnx
and $\mu$ and $\rho$ depend on $N$ and parameters of the harmonic oscillator in the continuum
\eqn
\rho=\frac{\omega^2 T^2}{2 (N-1)^2},\;\;\;
\mu=\frac{m (N-1)}{2\hbar T}.   \nn
\eqnx

This is the main modification compared to the free particle. With the first limit taken along the trajectory (\ref{hotraj}) the action (\ref{SN}) provides a positive representation for Minkowski path integral of a one-dimensional harmonic oscillator. 

Positivity of the restricted $P(z_i,z_i^*)$ is evident from the construction (\ref{SN}) while the normalizability can be seen by inspecting the eigenvalues of the real form $S_N({x_i,y_i})$. All but one eigenvalues are indeed positive. Due to the translational symmetry of the free particle problem, one eigenvalue is zero. The same eigenvalue becomes negative for the harmonic oscillator representation. Both of them can be dealt with in a standard manner.

Perhaps the most interesting example is that of a charged particle in a constant magnetic field, as it  does not admit positive representation
even after the Wick rotation. On the other hand the present approach offers a positive solution. It is well known since the time of Landau that the problem can be mapped into a one dimensional harmonic oscillator with the shifted centre of oscillations. Therefore we need only to rephrase
this equivalence in the language of propagators. To this end rewrite the Feynman kernel in the gauge used by Landau, $\vec{A}=(0,B,0)$,
 \eqn
K_{LG}(x_b,y_b, T; x_a,y_a, 0)=\exp{ \left\{ \frac{i m}{2 \hbar}  \left(  \frac{\omega}{2} \cot{\frac{\omega T}{2}}\left(\Delta x^2+\Delta y^2\right) + \omega (  x_a + x_b) \Delta y  \right)    \right\}, }  \nn
\label{KLG}
\eqnx
and split it as follows $(\Delta x = x_b-x_a, \Delta y = y_b - y_a)$
\eqn
K_{LG}=\underbrace{\exp{ \left\{ \frac{i}{\hbar} \frac{m}{2} \left( \omega (  x_a + x_b) \Delta y  + \omega ct {\Delta y}^2\right)    \right\} }}_{\exp{\left(\frac{i}{\hbar} p_y (y_b-y_a)\right)}=\exp{\left(\frac{i}{\hbar}m \omega O_x (y_b-y_a)\right)}} 
\underbrace{\exp{ \left\{ \frac{i}{\hbar} \frac{m}{2} \left(  \frac{\omega}{2} \cot{\frac{\omega T}{2}}\left(\Delta x^2-\Delta y^2\right)  \right)    \right\}. } }_{K^{HO}_{O_x}}\label{hoox}  
\eqnx
This exactly corresponds to the Landau solution of the Schr\"{o}dinger equation by factoring out the cyclic variable $y$.

To illustrate explicitly how the present proposal works in this case, consider the time dependent average position of a quantum particle subject to a boundary conditions
\eqn
\la \vec{x} \ra = \int d^2 x K(\vec{x}_b,\vec{x};T-t) \vec{x} K(\vec{x},\vec{x}_a;t) / K(\vec{x}_b,\vec{x}_a;T) = x^{cl}_{x_a,x_b,T} (t). \label{MFav}  \nn
\eqnx
Since the problem is gaussian the answer is given by the corresponding classical trajectory. Moreover due to the above equivalence 
the following reduction formulas hold
\eqn
\la x(t) \ra_{B}  = \la x(t) \ra_{O=O_x},\;\;\;\;\la y(t) \ra_{B}  = \la y(t) \ra_{O=O_y},  \label{rav}
 \nn
\eqnx
by which the "magnetic field averages" are expressed by the "shifted harmonic oscillator averages". And these can be obtained from the positive representation (\ref{SN}) trivially modified for the case of shifted oscillations.

In Fig.1 two classical trajectories of a charged particle in an external magnetic field are shown. They have the same corresponding initial and final points and differ by the total time $T$ of the transition from $\vec{x}_a$ to $\vec{x}_b$. Dots represent the averages calculated for the discretized problem using the above harmonic oscillator equivalence and positive representation of the shifted harmonic oscillator kernel. The first limit (\ref{hotraj}) has already been taken.
\begin{figure}[h]
\begin{center}
\includegraphics[width=7cm]{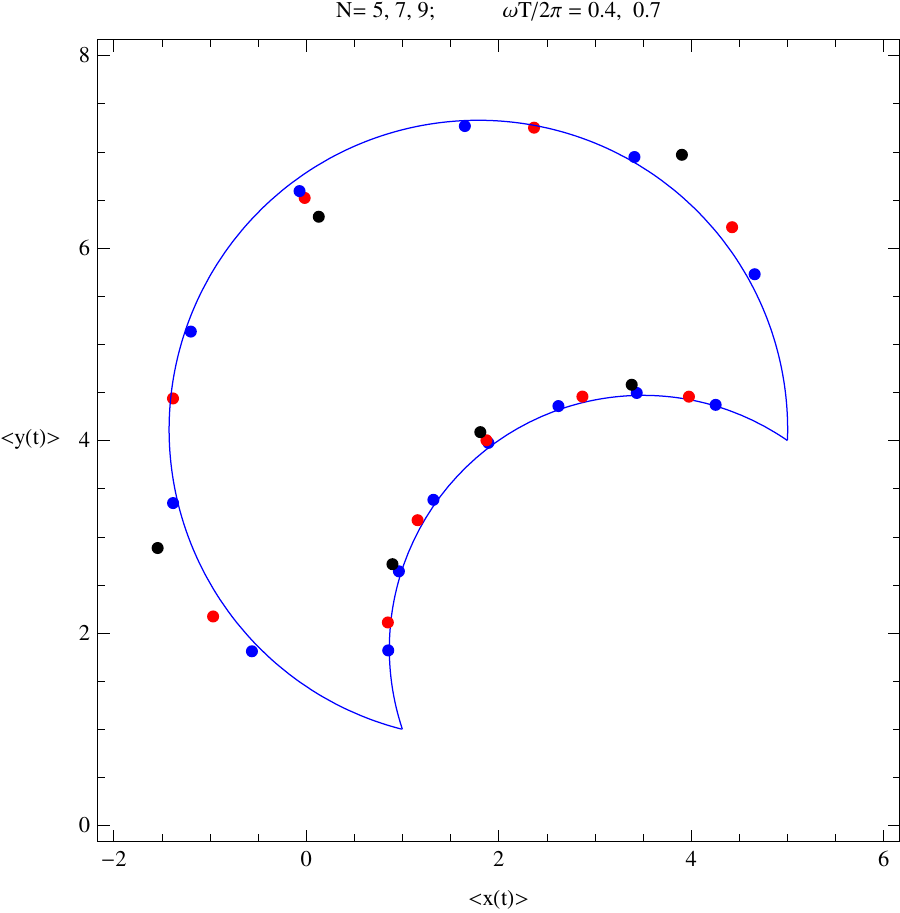}
\end{center}
\vskip-4mm \caption{Two exact trajectories of a particle in an external magnetic field compared with the quantum, discretized  averages obtained from the positive representation discussed in the text.} \label{fig:f1}
\end{figure}

The fast convergence (notice the very rough discretization) to the continuum is the tribute to the old and well known Feynman prescription. The new element however is that the averages were obtained from the positive representation of the harmonic motion directly in the Minkowski time. 

\noindent {\em Summary and outlook.}

Difficulties with stochastic solutions of the complex Langevin equations can be alleviated by the direct construction of pairs of corresponding, complex and positive, densities. 
This was recently done in Ref.\cite{Ja1} for the one degree of freedom case model thereby generalizing the existing solution of the gaussian model to an arbitrary complex slope. A particular nonlinear problem with a quartic coupling was also solved.

Subsequently the gaussian solution was generalized to many variables and used to construct the positive representation 
for gaussian path integrals directly in the Minkowski time \cite{Ja}. For the infinite number of degrees of freedom existence of the continuum limit is not trivial and requires an novel tuning of parameters. 

The method is then successfully applied to the three textbook quantum mechanical problems: 
a free particle, a harmonic oscillator and a particle in a constant, external magnetic field. The latter is the simplest prototype of a Wilson loop and until now did not have a positive representation even after the Wick rotation. 

Even in the context of above simple cases, many questions remain open: how fast is the first limit achieved in practice, how this depends on discretization, is there a more optimal way to combine the first limit with the continuum limit, etc.

Clearly one would like to generalize the present scheme to more physical, nonlinear systems. In particular  an interesting mathematical problem arises. Namely to what extent the sum rule (\ref{Pro}), together with positivity and normalizability conditions,
can determine $P$ from a complex weight $\rho$. The nonlinear example solved in Ref.\cite{Ja1} shows, that the new structure is not  restricted to the gaussian cases only. 

A host of further problems and applications suggests itself: generalization to compact integrals, nonlinear and nonabelian couplings, fermionic integrals, as well as extensions to the field theory are only few examples. 
We are looking forward to study some of them.

Finally, an intriguing interpretation may be enjoyed.  
The positivity is achieved by duplicating the number of variables. More precisely, positive Minkowski amplitudes would result 
if one specifies initial/final values for {\em all} of these variables. On the other hand 
the standard, complex quantum amplitudes emerge upon suitable integrations over half of above variables with the usual boundary  conditions. All this resembles to some extent the celebrated history of hidden variables. At the same time we strongly emphasize that none of the sacred principles of quantum mechanics is violated. In particular the quantum interference of standard amplitudes is not hampered in any way. Nevertheless, some new structure has been exposed and it remains to be seen if it is of practical interest only, or if it is more fundamental.  


\noindent{\em Acknowledgements}  
  
I would like to thank Erhard Seiler for many interesting discussions.
Existence of positive representations for complex, gaussian and more general, densities has been discussed in \cite{Sa,We}. The projection relation employed in \cite{Sa} bears some similarity to the condition (\ref{Pro}).

This work is supported in part by the NCN grant: UMO-2016/21/B/ST2/01492.

\end{document}